\documentclass[
 reprint,
 superscriptaddress,
 amsmath,amssymb,
 aps,
 prb,
]{revtex4-2}

\usepackage{graphicx}
\usepackage{dcolumn}
\usepackage{mathtools} %
\usepackage{bm}
\usepackage{braket}
\usepackage{physics}
\usepackage{appendix}
\usepackage{ulem}
\newcommand{\ve}[1]{{\mbox{\boldmath $#1$}}}

\usepackage[unicode=true]{hyperref}
\usepackage[dvipsnames]{xcolor} 

\hypersetup{
  colorlinks = true, 
  linkcolor = blue, 
  filecolor = blue, 
  urlcolor = blue, 
  citecolor = blue 
}
\makeatletter
\providecommand{\href}[2]{#2}
\providecommand{\url}[1]{#1}
\makeatother

\begin{document}


\title{Diagnosing energy gap in quantum spin liquids via polarization amplitude}

\author{Takayuki Yokoyama}
\affiliation{Quantum Matter Program, Graduate School of Advanced Science and Engineering, Hiroshima University,
Higashihiroshima, Hiroshima 739-8530, Japan}
\email[]{yokotaka308@hiroshima-u.ac.jp}

\author{Yasuhiro Tada}
\affiliation{Quantum Matter Program, Graduate School of Advanced Science and Engineering, Hiroshima University,
Higashihiroshima, Hiroshima 739-8530, Japan}
\affiliation{Institute for Solid State Physics, University of Tokyo, Kashiwa 277-8581, Japan}

\begin{abstract}
Identifying whether a many-body ground state is gapped or gapless is a fundamental yet challenging problem, especially in quantum spin liquids. In this work, we develop a gap-diagnostic scheme based on the polarization amplitude defined via twist operator, evaluated within the infinite density-matrix renormalization group (iDMRG) framework. As a benchmark, an analysis of the spin-$1/2$ XXZ chain demonstrates that the polarization amplitude clearly distinguishes the gapless Tomonaga–Luttinger liquid from the gapped Néel phase. We then extend this framework to infinite cylinders of the spin-$1/2$ XY–$J_\chi$ model on the square lattice. We find that the polarization amplitude sharply detects the transition between the gapless XY phase and the gapped chiral spin liquid phase.
These results show that polarization amplitudes provide a strong energy gap diagnostic in two-dimensional frustrated quantum magnets, including quantum spin liquids.
\end{abstract}

\maketitle

\section{Introduction}
In quantum many-body systems, one of the most basic questions is whether the ground state is gapped or gapless. This distinction strongly affects low-temperature properties and the nature of quasi-particle excitations~\cite{Wen_2007, Zeng_2019}.
It is especially important in strongly correlated systems such as frustrated magnets, where many different quantum phases can appear. In such systems including quantum spin liquids, it is often difficult to characterize the phase by a simple local order parameter~\cite{Savary_2017}. For this reason, determining the presence or absence of an excitation gap is a key step toward identifying the phase and understanding its low-energy physics.

However, a controlled numerical determination of excitation gaps remains highly nontrivial, given that it is an intrinsically quantitative issue. Especially in higher dimensions, the exponential growth of the Hilbert space limits calculations to finite systems. As a result, excitation gaps must be extracted from finite-size data, but the accessible sizes are often too limited for a reliable extrapolation to the thermodynamic limit. The situation is further complicated in strongly frustrated lattices, such as triangular and Kagome lattices, where computationally accessible system sizes are particularly limited~\cite{PRB.87.060405, Sun2024, PRX.15.011047}. 

Motivated by these issues, an alternative approach has been developed based on the twist operator as a qualitative indicator of gapless versus gapped behavior. For one-dimensional periodic systems, the polarization defined as the expectation value of the twist operator shows that its amplitude behaves as a quantized order parameter distinguishing gapless systems from gapped systems~\cite{PRL.80.1800, PRL.82.370, AligiaOrtiz1999,OrtizAligia2000,PRB.62.1666, Resta_2002}. This claim is investigated and extended in numerous analytical and numerical studies not only in itinerant particle systems~\cite{PRB.64.115202, PRB.65.153110,WatanabeOshikawa2018,PRB.99.085126, PRB.61.7883, PRB.107.075153} but also in quantum spin chains~\cite{PRL.89.077204, PhysRevB.97.165133, FuruyaNakamura2019}. A general proof for a part of this statement was given by Tasaki, who rigorously showed that, in one-dimensional U(1)-symmetric uniquely gapped systems, the absolute value of the ground-state expectation value of the twist operator converges to unity in the thermodynamic limit~\cite{Tasaki2018,Tasaki2022inbook,Tasaki2023}. The proof is based on the local operator formalism for infinite systems, where the thermodynamic limit is taken from the beginning. For sufficiently large finite systems, a concise proof is also available~\cite{Tada2024}, and in the presence of the enlarged SO(3) symmetry, even stronger statements have been proposed~\cite{Su2025}. On the other hand, for gapless finite systems, although it is not obvious that the ground state expectation value of the twist operator vanishes in the thermodynamic limit, it was shown that this property indeed holds in general~\cite{Tada2025}.  
However, the vanishing amplitude of the twist operator in a gapless infinite system has not been fully proven in a mathematically complete sense and thus its quantization remains unproven in the most general setting. Furthermore, it is known that the amplitude of the twist operator can be non-universal in higher dimensions and extension to two-dimensional systems is not straightforward~\cite{WatanabeOshikawa2018,Tada2025}. This poses a fundamental question regarding the usefulness of the twist operator for describing ordered states that emerge in higher-dimensional systems such as quantum spin liquids where diagnosis of an energy gap is a highly non-trivial problem~\cite{Savary_2017}.

In this work, we investigate whether the twist-operator-based energy-gap diagnosis in one dimension can be extended to quasi-one-dimensional infinite systems represented by cylinder geometry. Specifically, we consider the twist operator associated with the U(1) spin rotation symmetry, and evaluate its expectation value using infinite-DMRG (iDMRG) which has been widely used in the study of one-dimensional systems and also quantum spin liquids in cylinder geometry~\cite{PRL.69.2863, SCHOLLWOCK201196, McCulloch_2007, Tenpy, PRX.7.031020, PRB.97.075126}. Such cylinder geometries arise naturally within DMRG-based approaches, where two-dimensional systems are commonly approximated as infinitely long cylinders with finite circumference. While the twist operator and iDMRG have both been extensively studied, a systematic evaluation of the polarization amplitude on infinite cylinders within the iDMRG framework has not been carried out so far.
In contrast to finite-size approaches, the iDMRG formulation directly represents the thermodynamic limit along the cylinder axis and allows the twist operator to be applied on regions of increasing length without changing the underlying ground state. This provides a conceptually clean setting to examine whether the polarization amplitude can serve as a practical gap diagnostic in quasi-one-dimensional geometries relevant to two-dimensional quantum magnets.
We first discuss the $S=1/2$ XXZ chain as a benchmark and demonstrate that the behavior of the polarization amplitude is consistent with a spin gap within iDMRG. We then apply the method to a two-dimensional quantum spin model on infinitely long cylinders and show that the polarization amplitude can clearly distinguish a gapless XY phase from a gapped quantum spin liquid phase. 
Our results demonstrate the practical feasibility of qualitatively diagnosing gapless-gapped transitions in quasi-two-dimensional systems in a controlled infinite-system setting.

\section{\label{sec:II}Polarization amplitude for infinite chain}
The primary purpose of this work is to study cylinder systems with possible extrapolation to two dimensions in mind based on the twist operator. For this problem, we first discuss an infinite one-dimensional chain as a benchmark for the validity of iDMRG calculations of the twist operator.
Throughout this study, we consider systems with U(1) and translation symmetries.

\subsection{\label{sec:def_tw}Quantization of polarization amplitude}
\begin{figure}[tb]
  \centering
  \includegraphics[width=1.0\columnwidth]{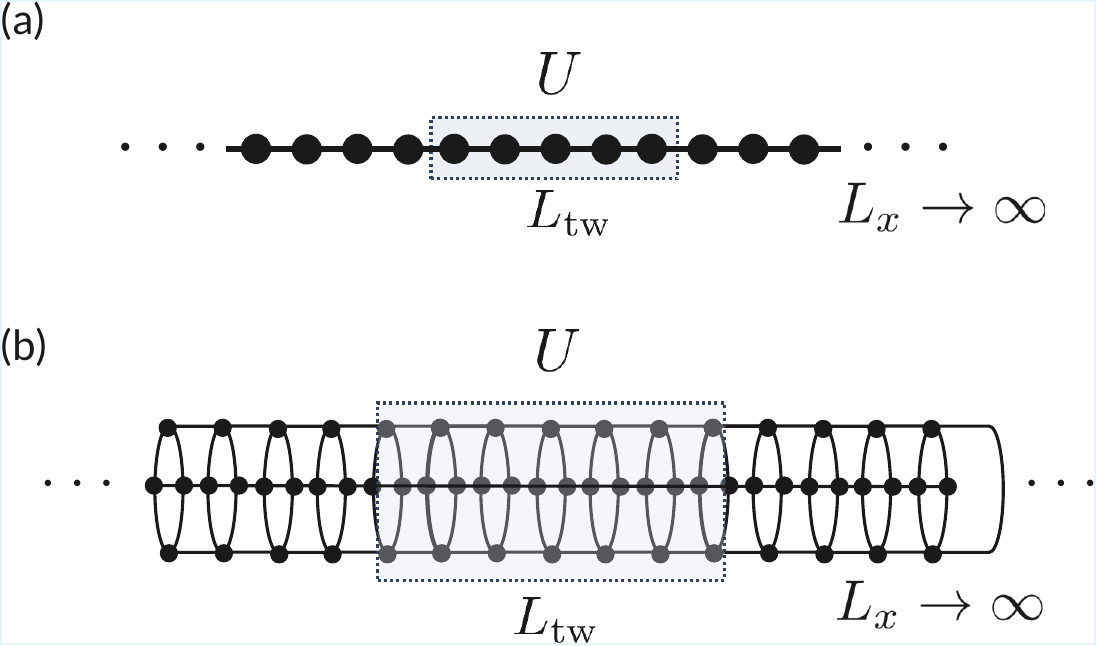}
  \caption{Schematic illustration of the spatial region on which the twist operator acts. (a) For an infinite one-dimensional chain in the thermodynamic limit ($L_x \to \infty$), the twist operator $U$ is applied to a finite region of length $L_{\mathrm{tw}}$. (b) For an infinite cylinder with circumference $L_y$ (here $L_y = 4$) and infinite axial length ($L_x \to \infty$), the twist operator $U$ is applied to a finite region of length $L_{\mathrm{tw}}$ along the axial direction.}
  \label{fig:geometry}
\end{figure}

In this work, we employ the twist operator as a ground-state indicator to diagnose whether a system is gapped or gapless~\cite{PRL.80.1800, PRL.82.370, AligiaOrtiz1999,OrtizAligia2000,PRB.62.1666, Resta_2002}. In a finite size system with the periodic boundary condition, the twist operator is defined as a nonlocal (large) U(1) gauge transformation associated with a conserved U(1) charge, such as the $z$ component of spin. On the other hand, in an infinite system, the twist operator is local and acts on a region of the system~\cite{Tasaki2018,Tasaki2022inbook,Tasaki2023}. Below we consider one-dimensional infinite systems and denote the spin operators on site $j$ by $S_j^z$. As schematically illustrated in Fig.~\ref{fig:geometry}(a), the twist operator is defined as
\begin{align}
  U
  &= \exp\!\left(
      i\frac{2\pi}{L_{\mathrm{tw}}}\, \sum_{j=1}^{L_{\mathrm{tw}}} j \, S^z_j
    \right),
  \label{eq:u_spin}
\end{align}
where $L_{\mathrm{tw}}$ is the length of the region on which the twist is applied~\cite{Tasaki2018,Tasaki2022inbook,Tasaki2023}. 
For finite-size systems, $L_{\mathrm{tw}}$ is often chosen as the system length $L$, but in iDMRG calculations, the system size is infinite from the beginning. 
The expectation value of the twist operator is closely tied to the presence or absence of a spin gap. The spin gap is defined as
\begin{equation}
  \Delta E(M)
  \equiv E(M+1)+E(M-1)-2E(M) ,
\end{equation}
where the conserved quantity is the total magnetization $M=\sum_j S_j^z$ of the super-unit cell within the iDMRG calculation. (One can also consider a spin gap for additional magnetization  localized in a finite segment~\cite{PRX.7.031020}.) For a spin-$S$ system, we define the corresponding filling $\nu_{\mathrm{s}}=(S-m)=p/q$ where $m$ is the magnetization per physical unit cell and $p,q$ are coprime integers. For $M=0$, the exponent is $q=2$ when $S$ is a half-integer and $q=1$ when $S$ is an integer. 
For a one-dimensional system at filling $\nu_{\mathrm{s}}$, we introduce a working ansatz
\begin{equation}
  |z^q|
  \coloneqq \lim_{L_{\rm tw} \to \infty}
  \left|
    \bigl\langle U^q \bigr\rangle
  \right|
  =
  \begin{cases}
    1, & (\Delta E>0) \\[2pt]
    0, & (\Delta E=0).
  \end{cases}
  \label{eq:quantized_polarization_spin}
\end{equation}
A key advantage of this approach is that the diagnosis can be carried out using only the ground-state wave function, without separately computing excited states. An intuitive understanding of the quantization of $|z^q|$ is that the twist for nearby spins with the small angle $O(1/L_{\rm tw})$ cannot create an excitation in a gapped system, while it changes the ground state to a low-energy excited state in a gapless system. Numerical estimation of an excitation gap is often difficult, because the gap size can be small and extrapolation to the thermodynamic limit may be subtle. On the other hand, for the twist operator, the extrapolation to $L_{\rm tw}\to\infty$ gives a quantized 0/1 value of $|z^q|$. In this work, we numerically evaluate $|z^q|$ and use its systematic behavior as an indicator of gapped versus gapless phases. 

We note that the strict quantization on the right-hand side of Eq.~\eqref{eq:quantized_polarization_spin} has been rigorously established only for large finite-size systems with periodic boundary conditions under certain conditions~\cite{Tada2024,Su2025,Tada2025}. For infinite systems, while it was proven that $|z^q|=1$ in uniquely gapped cases~\cite{Tasaki2018,Tasaki2022inbook,Tasaki2023}, it is not generally proven that $\ev{U^q}$ in a gapless system converges strictly to zero. Besides, it was not fully established whether or not the spin gap $\Delta E$ alone can completely determine behaviors of $|z^q|$. Namely, quantization of $|z^q|$ has not been proven for a system with a spin gap $\Delta E>0$ and a vanishing gap within the ground state $M$-sector~\cite{Tada2025}. (A related discussion is given in Appendix~\ref{sec:Hubbard}.) Nevertheless, the ansatz has been extensively used and repeatedly confirmed in numerical and analytical calculations for various one-dimensional finite systems under the periodic boundary condition~\cite{PRL.80.1800, PRL.82.370, AligiaOrtiz1999,OrtizAligia2000,PRB.62.1666, Resta_2002,PRB.64.115202, PRB.65.153110,WatanabeOshikawa2018,PRB.99.085126,PRL.89.077204, PhysRevB.97.165133, FuruyaNakamura2019}. We will numerically demonstrate that the ansatz indeed holds in infinite one-dimensional systems.

\subsection{Infinite density-matrix renormalization group}
\label{subsec:idmrg}

We obtain ground states using infinite density-matrix renormalization group (iDMRG). Since iDMRG directly targets the thermodynamic limit for one-dimensional chains and quasi one-dimensional systems like cylinder geometry, it allows us to avoid an explicit extrapolation of the system length. The absolute expectation value of the twist operator is evaluated by representing the exponential operator acting on a finite interval as a matrix-product operator (MPO) and contracting it with the resulting infinite matrix-product state (iMPS). Note that, in the conventional finite-size DMRG, the twist operator is usually defined on the whole system with the size $L_x$ and the ground states for each $L_x$ have to be calculated to obtain the expectation value $\ev{U^q}$ in the thermodynamic limit $L_x\to\infty$. In contrast, within iDMRG we can evaluate the corresponding thermodynamic-limit expectation value more directly by using the MPO and iMPS formalism only with a single infinite-size ground state, which is an advantage in numerical calculations.
By systematically increasing the operator length $L_{\mathrm{tw}}$, we can probe the convergence of $\ev{U^q}$ within the infinite-system framework. All calculations are performed with total $S^z$ conservation. The truncation error is kept below $10^{-7}$ for one-dimensional chains (Sec.\ref{sec:xxz}) and below $10^{-5}$ for quasi-one-dimensional cylinders (Sec.\ref{sec:Jchi}).

\subsection{Numerical benchmark for infinite XXZ chain}
\label{sec:xxz}
\begin{figure}[tb]
  \centering
  \includegraphics[width=1.0\columnwidth]{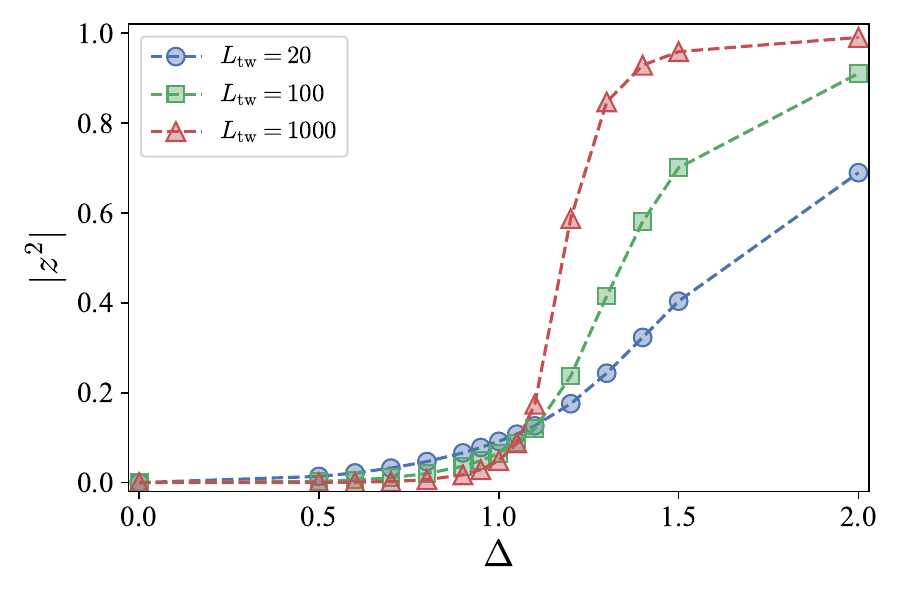}
  \caption{Spin polarization amplitude $|z^2|$ of the XXZ chain as a function of the anisotropy $\Delta$. Blue circles, green squares, and red triangles correspond to $L_{\mathrm{tw}}=20,\,100,\,1000$, respectively. Here $L_{\mathrm{tw}}$ denotes the region length on which the twist operator is applied and should not be confused with the system size.}
  \label{fig:xxz_U}
\end{figure}
We study the $S=1/2$ antiferromagnetic XXZ chain with the nearest neighbor interactions. The model is described by
\begin{equation}
H_{\text{XXZ}} = J \sum_{j} \left(S_j^x S_{j+1}^x + S_j^y S_{j+1}^y
      + \Delta \, S_j^z S_{j+1}^z \right),
  \label{eq:ham_xxz_chain}
\end{equation}
where $S^{\alpha}$ ($\alpha = x,y,z$) is the spin operator and $J>0$ denotes the antiferromagnetic exchange coupling, and $\Delta$ parametrizes the exchange anisotropy of the $z$ component. In this work, we focus on the regime $0\leq \Delta$. The phase diagram of the ground state has been well established: for $\Delta\le 1$ it realizes the gapless Tomonaga-Luttinger liquid (TLL), whereas for $\Delta>1$ it does the gapped Néel order phase~\cite{text_Giamarchi}. Here, we examine whether the twist-operator quantity $|z^2|$ can detect the quantum phase transition between the TLL and the Néel phases within iDMRG as previously demonstrated by other numerical methods.

Figure~\ref{fig:xxz_U} shows $|z^2|$ evaluated by applying the twist operators on a finite region of length $L_{\mathrm{tw}}$ embedded in an iMPS. We find that $|z^2|$ changes from zero to a nonzero value around $\Delta\simeq 1.0$, signaling the known transition between the TLL and the Néel phases. Moreover, as $L_{\mathrm{tw}}$ is increased, $|z^2|$ becomes smaller in the gapless TLL phase, while it grows in the gapped Néel phase and exhibits a tendency to approach $|z^2|\to 1$. This behavior suggests that the quantization of the spin polarization amplitude $|z^2|$, known for finite one-dimensional systems with periodic boundary conditions, also emerges in the iMPS-based evaluation.
It is worth emphasizing that this is a natural but non-trivial result, since the quantization of $|z^2|$ in the limit $L_{\rm tw}\to\infty$ has not been analytically proven for general infinite systems. 
We also obtain similar results in the Hubbard model where there are two U(1) symmetries associated with the charge and spin degrees of freedom as discussed in Appendix~\ref{sec:Hubbard}. These results indicate that the polarization amplitude is capable of detecting the gapped–gapless transition not only in spin systems but also in spinful fermion systems.

We point out that iDMRG calculations require care regarding the bond-dimension dependence of $|z^2|$. In a gapless phase, the correlation length should diverge, but in iDMRG a finite bond dimension $D$ induces a finite energy scale (an effective gap) $\Delta_{D}$, leading to a finite correlation length $\xi_D \propto 1/\Delta_D$. Therefore, the region size $L_{\mathrm{tw}}$ should be chosen such that $L_{\mathrm{tw}} < \xi_D$. We discuss this point in detail in Appendix~\ref{app:bond_dependence}.

\section{Polarization amplitude in Quantum spin liquid}
\label{sec:Jchi}

\begin{figure}[tb]
  \centering
  \includegraphics[width=1.0\columnwidth]{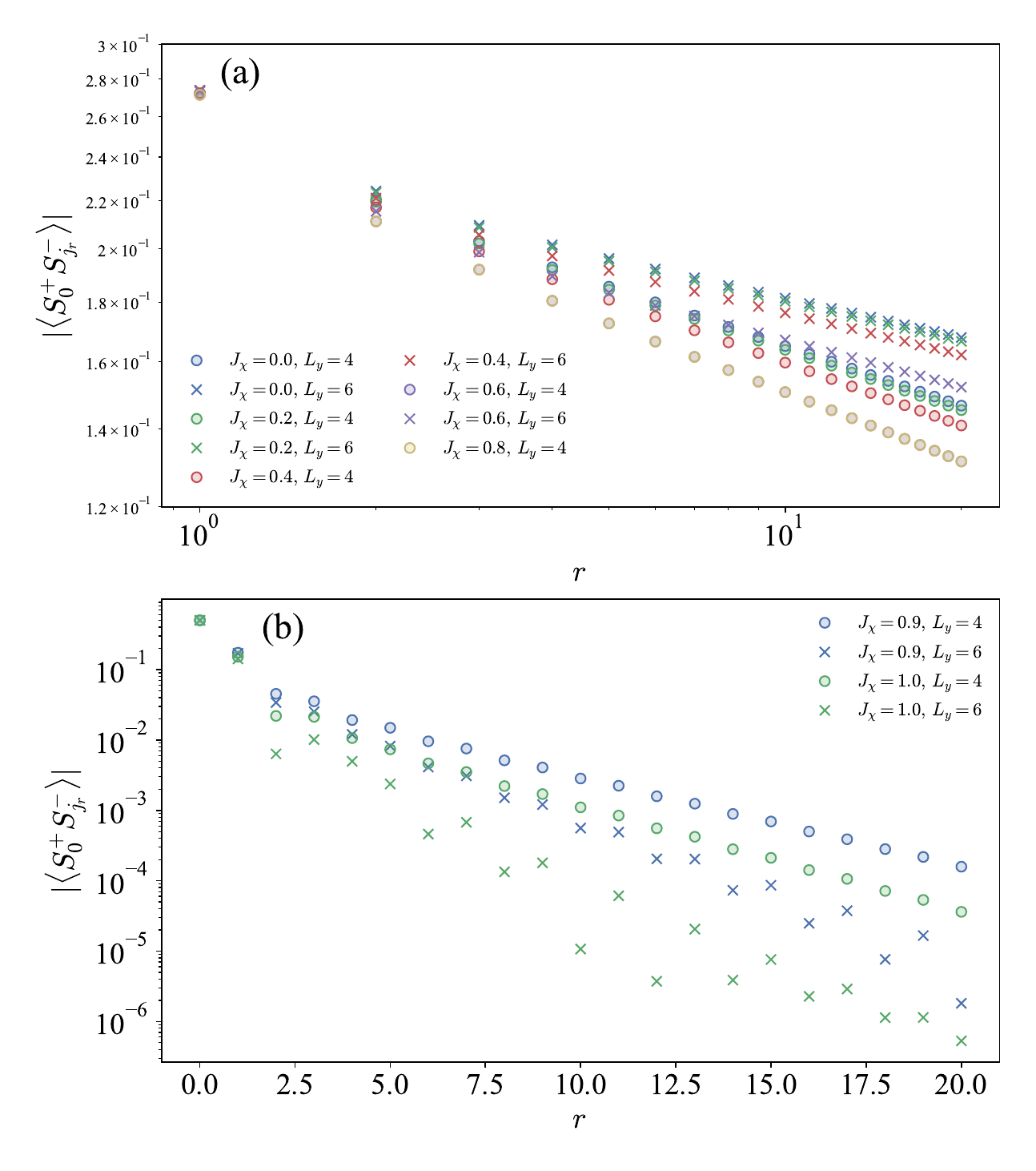}
  \caption{In-plane spin correlations in the square-lattice XY--$J_\chi$ model on infinite cylinders. We plot the equal-time correlation function $\bigl|\langle S^{+}_{0} S^{-}_{j_r}\rangle\bigr|$ as a function of the distance $r$ along the cylinder axis for circumferences $L_y=4$ (circles) and $L_y=6$ (crosses). (a) XY phase ($J_\chi=0.0$--$0.8$): the correlations show a slow, approximately algebraic decay, consistent with quasi-long-range order on cylinders. (b) Chiral spin liquid phase ($J_\chi=0.9,\,1.0$): the correlations decay much more rapidly over distance, indicating a finite correlation length and consistent with a non-zero gap.}

  \label{fig:Jchi_XY_neelcorr}
\end{figure}

To examine whether the method can be applied to two-dimensional quantum spin liquids, we study the $S=1/2$ XY--$J_\chi$ model on the square lattice. This model combines an in-plane (XY) exchange interaction with a scalar spin-chirality interaction defined on triangular plaquettes. The Hamiltonian is
\begin{equation}
  H_{\text{XY}J_\chi}
  = J \sum_{\langle i,j\rangle}
      \left(S_i^x S_j^x + S_i^y S_j^y\right)
    + J_\chi \sum_{\langle i,j,k\rangle_\triangle}
      \ve{S}_i \cdot \left(\ve{S}_j \times \ve{S}_k \right),
  \label{eq:ham_xy_jchi}
\end{equation}
where $J=1$ is the energy unit and $J_\chi\ge 0$, and $\langle i,j,k\rangle_\triangle$ denotes an ordered triplet of sites on a triangular plaquette taken in the clockwise direction. The $J_\chi$ term explicitly breaks time-reversal symmetry and can induce a gapped chiral spin-liquid phase on the square lattice. For the isotropic Heisenberg-$J_\chi$ model, it is known that a phase transition occurs around $J_\chi \simeq 0.8$ from a gapless SU(2) Néel phase to a chiral spin-liquid phase~\cite{PRB.105.155104, PRB.110.224404}. Therefore, we expect that the present XY-$J_\chi$ model also exhibits a similar phase transition. In this study, we investigate the system defined on an infinite cylinder with the circumference $L_y$ with use of iDMRG (Fig.~\ref{fig:geometry}(b)). Such a quantum phase transition would take place even for the quasi-one-dimensional cylinder if $L_y$ is sufficiently large. Numerically, the chiral spin liquid phase in the Heisenberg-$J_{\chi}$ model is stable even for a relatively small $L_y$~\cite{PRB.105.155104, PRB.110.224404}, since an energy gap induces a short correlation length.

\begin{figure*}[tb]
  \centering
  \includegraphics[width=\textwidth]{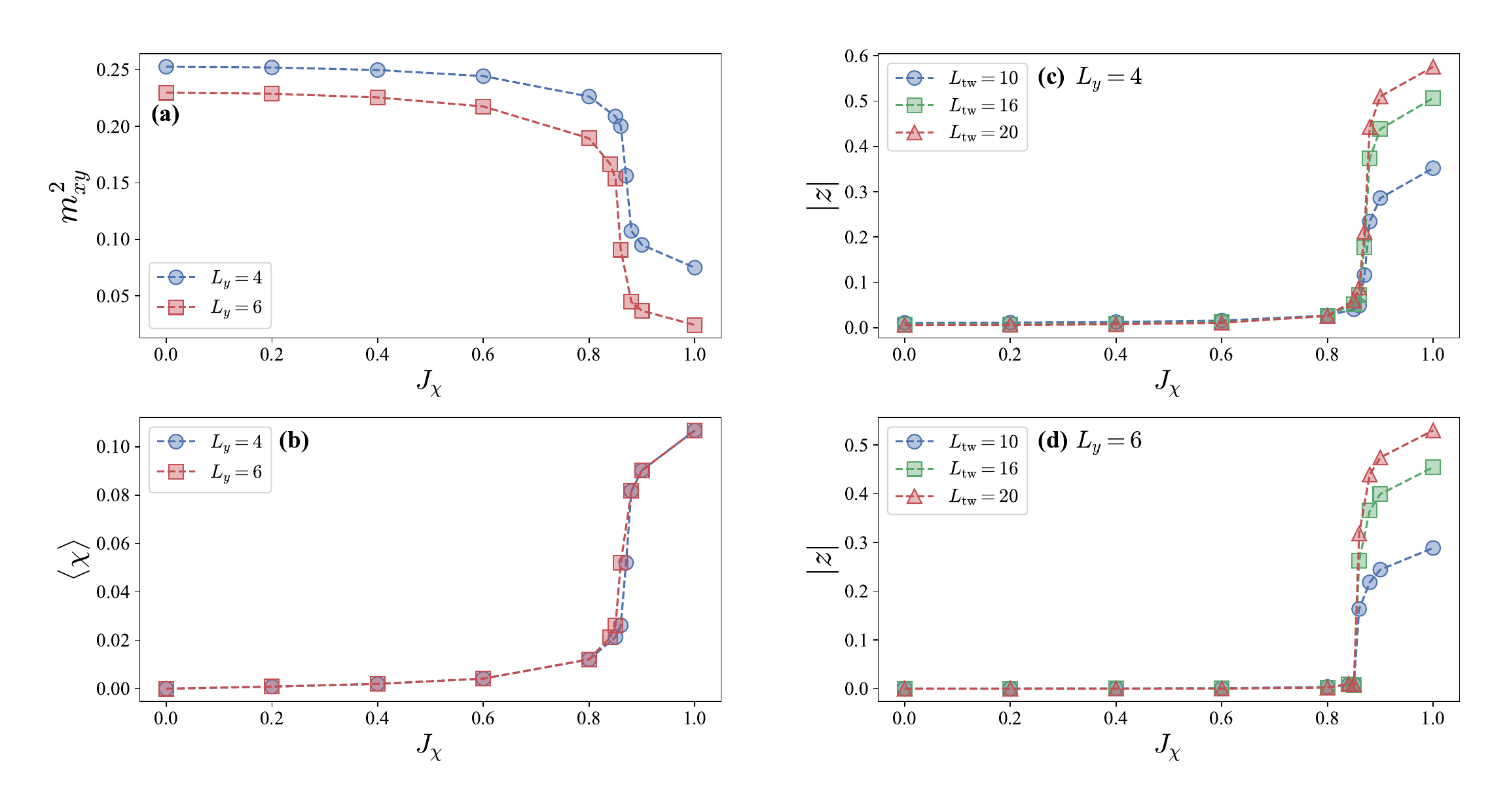}
  \caption{Ground-state diagnostics for the square-lattice $S=1/2$ XY--$J_\chi$ model on infinite cylinders. (a) In-plane Néel correlation $m_{xy}^2$ within the $L_y\times L_y$ region (Eq.\eqref{eq:neel}) as a function of $J_\chi$ for $L_y=4$ and $6$. (c) Scalar spin chirality $\langle \chi\rangle$ (Eq.~\eqref{eq:chiral}) as a function of $J_\chi$ for $L_y=4$ and $L_y = 6$, where. (a,b) Polarization amplitude $|z|$ as a function of $J_\chi$ for (a) circumference $L_y=4$ and (b) $L_y=6$. Different symbols correspond to different twist lengths $L_{\mathrm{tw}}$ (length along the cylinder axis) on which the twist operator is applied. The polarization amplitude changes from nearly zero to a finite value across the transition.}

  \label{fig:Jchi}
\end{figure*}

We first discuss the phase transition based on the conventional local order parameters, although such a quantity does not exist for general spin liquids.
To examine the presence or absence of magnetic order, we compute the Néel correlation function. 
On a cylinder with a finite circumference, the XY phase does not exhibit a true long-range order but instead is expected to realize a quasi-long-range-ordered state with algebraically decaying correlations. In contrast, once the magnetic order is suppressed and the system enters the expected gapped chiral spin liquid phase, the magnetic correlations will decay exponentially. Figure~\ref{fig:Jchi_XY_neelcorr} shows the in-plane correlation function $\langle S^{+}_{0} S^{-}_{j_r} \rangle$ for cylinders with circumferences $L_y=4$ and $6$, where $r$ is the distance between the origin (0,0) and $j_r=(r,0)$ along the cylinder direction. In the range $J_\chi=0.0 \sim 0.8$, the correlations exhibit an approximately algebraic decay, consistent with the quasi-long-range order expected for the XY phase on a cylinder. In contrast, for $J_\chi=0.9$ and $1.0$ we observe an exponential (or even faster) decay, indicating a finite correlation length. These results suggest that increasing $J_\chi$ strongly suppresses magnetic correlations and drives the system into a distinct gapped quantum phase.

To quantify the XY (in-plane Néel order) phase, we also evaluate the sum of in-plane spin correlations within an $L_y\times L_y$ region,
\begin{equation}
  m^2_{xy} = \frac{1}{L_y^2} \sum_{i,j} e^{i \mathbf{k} \cdot \mathbf{r}_{ij}}\ev{S^x_i S^x_j + S^y_i S^y_j },
  \label{eq:neel}
\end{equation}
where $\mathbf{k}=(\pi,\pi)$ and the summation is restricted to the region.
For square-lattice long cylinders, extrapolating $m_{xy}^2$ to the thermodynamic limit is empirically known to provide a good estimate of the squared Néel order parameter \cite{PRL.113.027201, PRB.105.155104}.
Figure~\ref{fig:Jchi}(a) shows the results for $m^2_{xy}$, for two different circumferences $L_y=4, 6$. We find that the in-plane Néel order is suppressed around $J_\chi \sim 0.8$ and it approaches zero as the circumference increases. This behavior suggests a quantum phase transition driven by the $J_\chi$-term from the Néel ordered phase to a quantum state without a conventional magnetic order in the large $L_y$ limit.

\begin{figure*}[tb]
  \centering
  \includegraphics[width=1.0\textwidth]{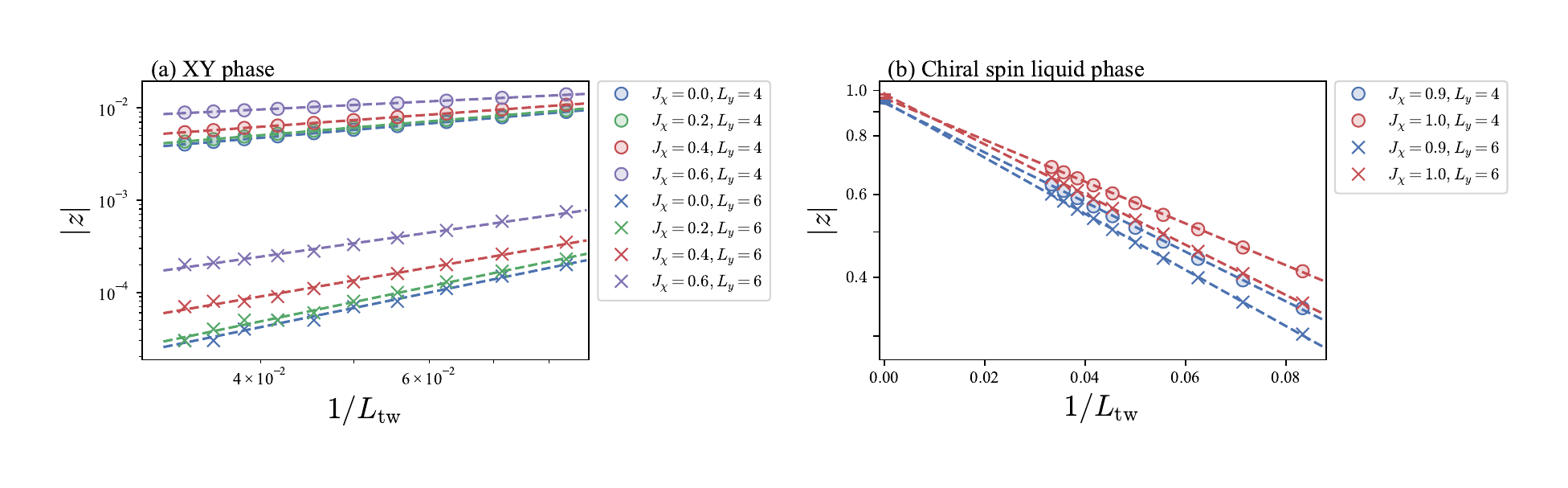}
  \caption{Scaling of the spin polarization amplitude with the inverse twist length in the square-lattice XY--$J_\chi$ model on infinite cylinders. The vertical axis shows the spin polarization amplitude $|z|$, while the horizontal axis represents the twist length $L_{\mathrm{tw}}$. Circle and cross symbols correspond to cylinders with circumferences $L_y = 4$ and $L_y = 6$, respectively. Within each set, the curves are ordered from bottom to top as $J_\chi = 0.0, 0.2, 0.4$, and $0.6$. (a) XY phase ($J_\chi = 0.0$--$0.6$): $|z|$ shows as a power-law decay and vanishes in the $L_{\mathrm{tw}} \to \infty$ limit, with dashed lines indicating power-law fits of the form $|z| \propto 1/L_{\mathrm{tw}}^{\alpha}$. (b) Chiral spin-liquid phase ($J_\chi = 0.9,\,1.0$): $|z|$ extrapolates toward values close to unity as $L_{\mathrm{tw}} \to \infty$, consistent with a gapped chiral spin-liquid regime. Dashed lines indicate exponential fits of the form $|z|= |z_{\infty}| e^{-\lambda/L_{\mathrm{tw}}}$, where $|z_{\infty}|,\lambda$ are fitting parameters.
}
  \label{fig:Jchi_XY_polarization_fitting}
\end{figure*}
To characterize the non-magnetic state at large $J_{\chi}$, we evaluate the scalar chirality on an elementary triangular plaquette,
\begin{equation}\label{eq:chiral}
  \ev{\chi_{ijk}} = 
      \ev{\ve{S}_i \cdot \left(\ve{S}_j \times \ve{S}_k \right)}.
\end{equation}
$\ev{\chi_{ijk}}$ is not a sharp order parameter which qualitatively characterizes the quantum phase transition, since the Hamiltonian does not possess time-reversal symmetry and hence $\ev{\chi_{ijk}}$ can be non-zero even in the XY phase when $J_{\chi}\neq0$. From the behaviors of $\ev{\chi_{ijk}}$, however, we can quantify the degree of time-reversal symmetry breaking in the ground state. Indeed, as shown in Fig.~\ref{fig:Jchi}(b), the results for the spin scalar chirality $\ev{\chi}$ indicate that time-reversal symmetry is nearly kept in the XY phase and is strongly broken in the large-$J_\chi$ regime, which is consistent with a key feature of chiral spin liquids. These results support the presence of the chiral spin liquid state and the quantum phase transition from the XY phase in the XY-$J_\chi$ model on the square lattice cylinder. Our identification of the chiral spin liquid phase is consistent with previous studies~\cite{PRB.105.155104, PRB.110.224404}. We expect that the chiral spin liquid state is stable even for the large $L_y$ limit, because it is a gapped phase for which the circumference will exceed the correlation length. In this sense, the chiral spin liquid phase in the finite-$L_y$ cylinders could be regarded as an essentially two-dimensional quantum order, as widely believed in iDMRG calculations on general spin liquids.

Although the absence of conventional symmetry-breaking order can often be established numerically in candidate spin liquids, determining whether such a phase is truly gapped or gapless remains a highly delicate question. 
We now discuss the polarization amplitude in the XY--$J_{\chi}$ model on the infinite cylinder.
Taking the cylinder axis as the $x$ direction, we define the spin twist operator along the $x$-axis as
\begin{equation}
  U  = \exp\!\left(
      i\frac{2\pi}{L_{\mathrm{tw}}}
      \sum_{j\in\Lambda(L_{\mathrm{tw}})} x_j\, S_j^z
    \right),
  \label{eq:twist_xy_jchi}
\end{equation}
where $\Lambda(L_{\mathrm{tw}})$ denotes the set of sites in an $L_{\mathrm{tw}}\times L_y$ region and $x_j$ is the $x$ coordinate of site $j$ (Fig.~\ref{fig:geometry}(b)). For simplicity, we have used the same symbol $U$ as in Eq.~\eqref{eq:u_spin} for the twist operator. On a cylinder, the $L_y$ sites sharing the same $x$ coordinate can be viewed as belonging to the same unit cell; accordingly, the effective U(1) charge per unit cell is increased to $L_y(S-m)$. In particular, for even $L_y$ and $m=0$, we expect the spin polarization amplitude $|z|$ to be quantized, independent of whether $S$ is integer or half-integer. We note that Eq.~\eqref{eq:twist_xy_jchi} is defined for quasi-one-dimensional systems with $L_{\rm tw}\gg L_y$ and is distinct from a variant of the twist operator for isotropic two-dimensional systems~\cite{Tada2025}. Nevertheless, we will demonstrate that the quasi-one-dimensional twist operator can clearly diagnose an excitation gap. Figure~\ref{fig:Jchi}(c)(d) shows the result for $|z|$, which indicates the phase transition between the gapless XY phase and the gapped chiral spin liquid phase around $J_\chi \simeq 0.85$. In the gapless regime, $|z|$ takes a small value with only a weak dependence on the region size $L_{\rm tw}$, and it tends to approach zero as $L_{\rm tw}$ is increased. 

It should be noted that, while bosonization arguments are available for one-dimensional systems, extending them to isotropically coupled cylinder geometries is highly nontrivial, as the identification of low energy modes and their coupling to the twist operator becomes ambiguous. In this sense, numerical approaches such as iDMRG provide an indispensable route.
By contrast, in the gapped regime, $|z|$ increases monotonically with $L_{\rm tw}$ and shows a tendency to move toward values close to unity. These results suggest that, even in cylinder-geometry calculations, the polarization amplitude can be used to detect nontrivial two-dimensional phases such as a chiral spin liquid.

We also analyze the scaling of $|z|$ with the twist length $L_{\mathrm{tw}}$ to discuss the quantization of polarization amplitude. Previous studies in one-dimensional finite systems have shown that the polarization amplitude decays algebraically with system size $L$ in gapless systems~\cite{PhysRevB.97.165133,FuruyaNakamura2019,Tada2025}. In contrast, the precise system-size dependence of $|z^q|$ in gapped systems has been less explored numerically, although it was shown that $|z^q|\simeq 1+O(1/L)$~\cite{Tasaki2018,Tasaki2022inbook,Tasaki2023,Su2025,Tada2024,Tada2025} and was proposed that $|z^q|\simeq e^{-\lambda/L+O(1/L^2)}$ with a localization length $\lambda$ under some assumptions~\cite{PRL.82.370,PRB.62.1666}. Our analysis in the XY phase shows that the polarization amplitude $|z|$ exhibits a power-law decay as a function of $L_{\mathrm{tw}}$, and the corresponding exponent decreases as $J_\chi$ increases (Fig.\ref{fig:Jchi_XY_polarization_fitting}(a)). The algebraic behavior is consistent with the known results for other gapless systems such as the XXZ chain~\cite{PhysRevB.97.165133,FuruyaNakamura2019}. On the other hand, in the gapped chiral spin liquid, we find that $|z|$ is well described by an exponential approach to unity, $|z|\simeq e^{-\lambda/L_{\mathrm{tw}}}$ without additional $O(e^{{\rm const}/L_{\rm tw}^2})$ corrections in a wide range of $L_{\rm tw}$  (Fig.\ref{fig:Jchi_XY_polarization_fitting}(b)). This behavior indicates that the finite-size effect is rapidly suppressed as $L_{\mathrm{tw}}$ increases. Especially, by exponential-fitting of $L_{\mathrm{tw}}$ dependence of $|z|$ in the chiral spin liquid and extrapolating to $L_{\mathrm{tw}}\to\infty$, we obtain values close to unity (in the range $0.94\sim0.98$). This provides strong numerical evidence that the polarization amplitude is quantized in the spin liquids on the infinite cylinder.

\section{Summary}
\label{sec:summary}

In this work, we studied the polarization amplitude as diagnostic of gapped versus gapless excitations in the infinite one-dimensional chain and the infinite quasi-one-dimensional cylinder. Using iDMRG, we considered a finite region with length $L_{\mathrm{tw}}$ in an infinite system represented by an iMPS, and examined the convergence of $|z^q|$ as the twist length $L_{\mathrm{tw}}$ is increased. For $S=1/2$ XXZ chain, we found that the spin polarization amplitude $|z^2|$ correctly detects the transition between the gapless TLL phase and the gapped Néel ordered phase. 

We then applied the same framework to a two-dimensional quantum spin model on infinite cylinders, namely the $S=1/2$ XY-$J_\chi$ model on the square lattice. On the infinite cylinders with circumference $L_y$, the polarization amplitude $|z|$ changes clearly across a critical $J_\chi\simeq 0.85$. This behavior is consistent with the identification of the gapless XY phase and the gapped chiral spin liquid phase based on the in-plane Néel correlation and the spin scalar chirality. Although the polarization amplitude was originally formulated for one-dimensional systems with periodic boundary conditions, our results suggest that it is also useful in infinite one-dimensional chains and can be applied even to an infinite quasi-one-dimensional cylinder relevant to two-dimensional quantum magnets. The polarization amplitude provides a practical ground state criterion for gapped versus gapless excitations on infinite cylinders, and it opens a concrete route to tackle gap diagnostics in strongly frustrated two-dimensional magnets, including quantum spin liquids, within iDMRG calculations. Extending the present approach to genuinely two-dimensional systems beyond cylinder geometries remains an important direction for future work.

\begin{acknowledgements}
The iDMRG calculations were performed with the use of Tenpy liberally~\cite{Tenpy} (version 1.1.0). The computation in this work has been done using the Supercomputer Center, the Institute for Solid State Physics, the University of Tokyo. This work was supported by JST SPRING Grant No. JPMJSP2132 and JSPS KAKENHI Grant No. 22K03513.

\end{acknowledgements}

\appendix

\section{Polarization amplitude in infinite Fermi-Hubbard chain}
\label{sec:Hubbard}
\begin{figure}[t]
  \centering
  \includegraphics[width=1.0\columnwidth]{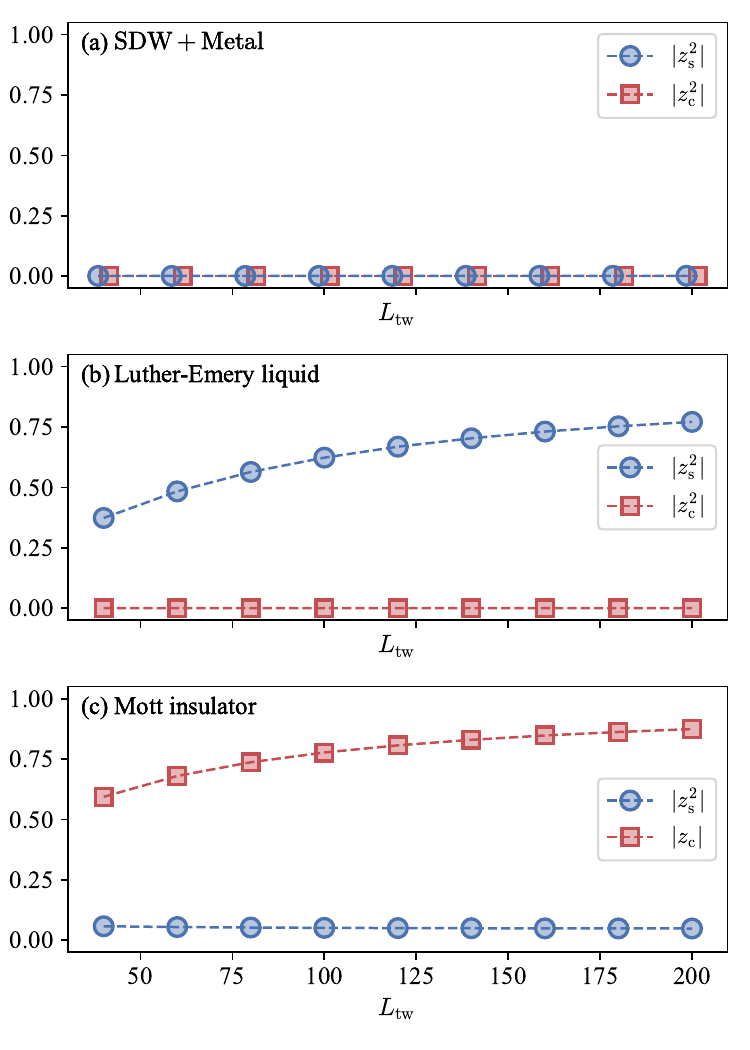}
  \caption{Spin and charge polarization amplitudes. The vertical axis shows $|z_{\mathrm{s}}^{q_{\mathrm{s}}}|$ (spin) and $|z_{\mathrm{c}}^{q_{\mathrm{c}}}|$ (charge), plotted as functions of the region length $L_{\mathrm{tw}}$ on which the twist operator is applied. Blue circles denote the spin polarization amplitude and red squares denote the charge polarization amplitude. (a) $U=2.0$ at $\nu_{\mathrm{c}}=1/2$, with $q_{\mathrm{s}}=2$ and $q_{\mathrm{c}}=2$. For visibility, the blue and red markers are slightly shifted horizontally, but they correspond to the same values of $L_{\mathrm{tw}}$. (b) $U=-2.0$ at $\nu_{\mathrm{c}}=1/2$, with $q_{\mathrm{s}}=2$ and $q_{\mathrm{c}}=2$. (c) $U=2.0$ at $\nu_{\mathrm{c}}=1.0$, with $q_{\mathrm{s}}=2$ and $q_{\mathrm{c}}=1$.}
  \label{fig:hubbard_1d}
\end{figure}
\begin{figure*}[t]
  \centering
  \includegraphics[width=\textwidth]{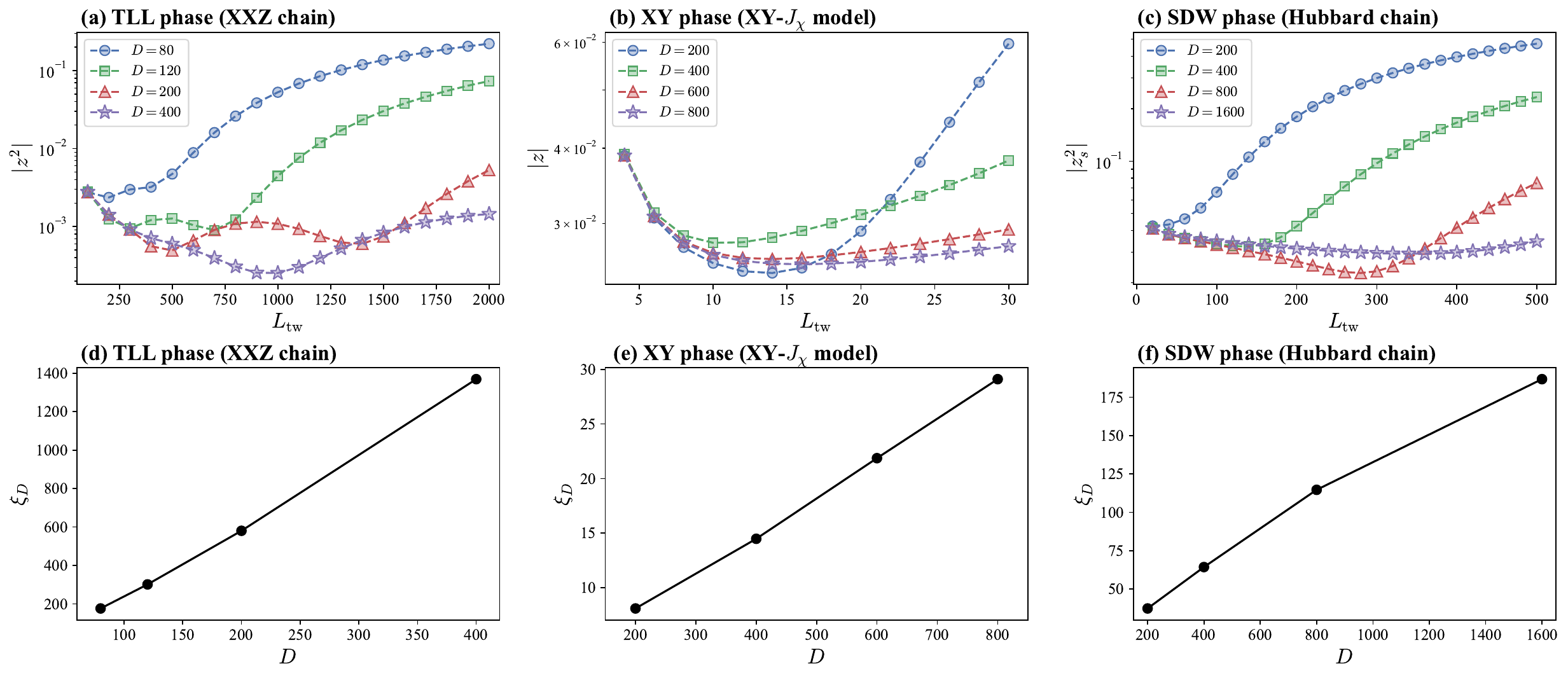}
  \caption{Dependence of polarization amplitude on bond-dimension gapless phases. 
(a--c) Result of spin polarization amplitude (a) the Tomonaga-Luttinger-liquid (TLL) phase of the $S=1/2$ XXZ chain at $\Delta = 0.5$, (b) the XY phase of the XY--$J_\chi$ model in square lattice on an infinite cylinder with circumference $L_y=4$ for $J_\chi =0.8$, and (b) the spin-density-wave (SDW) phase of the fermionic Hubbard chain for $U = 0.5$ at quarter filling $\nu=1/2$. (d--f) The corresponding finite-entanglement correlation length $\xi_D$ extracted from the transfer matrix eigenvalues ($\xi_D=-1/\ln|\lambda_2/\lambda_1|$), plotted versus bond-dimension $D$ for each case. 
In all gapless phases, $|z_{\mathrm{s}}^2|$ decreases toward zero only up to a crossover scale $L_{\mathrm{tw}}^{\ast}$, but turns upward for $L_{\mathrm{tw}}\gtrsim L_{\mathrm{tw}}^{\ast}$.}
  \label{fig:xi_dependice}
\end{figure*}
Here, we consider one-dimensional spinful fermions where there are two U(1) symmetries, namely the spin rotation symmetry and the particle number conservation. The twist operators for charge and spin are defined as
\begin{align}
  U_{\mathrm{c}}
  &= \exp\!\left(
      i\frac{2\pi}{L_{\mathrm{tw}}}\, \sum_{j=1}^{L_{\mathrm{tw}}} j n_j
    \right),
  \label{eq:uch_def}
  \\
  U_{\mathrm{s}}
  &= \exp\!\left(
      i\frac{2\pi}{L_{\mathrm{tw}}}\,\sum_{j=1}^{L_{\mathrm{tw}}} j S^z_j
    \right),
  \label{eq:usp_def}
\end{align}
where the particle number on site $j$ is denoted by $n_j$ and $L_{\mathrm{tw}}$ is the length of the region on which the twist is applied. Let $E(N, M)$ be the ground-state energy in the particle-number sector $N$ and the total spin sector $M$, we define the charge gap $\Delta E_{\mathrm{c}}(N, M)$ and spin gap $\Delta E_{\mathrm{s}}(N, M)$ by
\begin{align}
  \Delta E_{\mathrm{c}}(N, M)
  \equiv E(N+1, M)+E(N-1, M)-2E(N, M),  \\ \nonumber
  \Delta E_{\mathrm{s}}(N, M)
  \equiv E(N, M+1)+E(N, M-1)-2E(N, M). \\ 
\end{align}
For a one-dimensional system with a super-unit cell length $L$ at particle filling $\nu_{\mathrm{c}}=N/L=p_{\mathrm{c}}/q_{\mathrm{c}}$ and spin filling $\nu_{\mathrm{s}}=1/2-M/L=p_{\mathrm{s}}/q_{\mathrm{s}}$, 
we consider polarization amplitudes for each of the particle-number sector and the spin-sector, 
\begin{align}
  |z_{\alpha}^{q_{\alpha}}|
  \coloneqq
  \lim_{L_{\rm tw} \to \infty}
  \left|
    \bigl\langle U_{\alpha}^{q_{\alpha}} \bigr\rangle
  \right|,
  \quad (\alpha = \mathrm{c}, \mathrm{s}).
  \label{eq:def_z_alpha}
\end{align}
Similarly to the spin models in the main text, we introduce a working ansatz that,
for each sector $\alpha=\text{c, s}$, $|z^{q_\alpha}_{\alpha}|$ is quantized to unity in the presence of a finite gap $\Delta E_{\alpha} > 0$, while it vanishes in the gapless case with $\Delta E_{\alpha} = 0$. It should be noted that the quantization of the polarization amplitudes has not been fully proven, even if we assume that an infinite system is essentially the same as a finite large system with the periodic boundary condition. For example, quantization of $|z_{\rm c}^{q_{\rm c}}|$ has not been fully proven in a state where there is a non-zero charge gap ($\Delta E_{\rm c}>0$) and no spin gap ($\Delta E_{\rm s}=0$), although $|z_{\rm s}^{q_{\rm s}}|\to 0$ can be proven for this state~\cite{Tada2025}. One requires an additional (physically reasonable) assumption that there is no gapless charge density excitation within the $(N,M)$-sector to prove $|z^{q_{\rm c}}_{\rm c}|\to 1$. We will numerically demonstrate that the quantization ansatz of the polarization amplitudes indeed works in iDMRG calculations of spinful systems.

To be concrete, we consider the one-dimensional Fermi-Hubbard model.
The Hamiltonian is given by
\begin{equation}
  H_{\mathrm{Hub}}
  = -t \sum_{j,\sigma}
    \left(
      c_{j,\sigma}^\dagger c_{j+1,\sigma}
      + \mathrm{H.c.}
    \right)
    + U \sum_j n_{j,\uparrow} n_{j,\downarrow},
  \label{eq:ham_hubbard_chain}
\end{equation}
where $t=1$ is the nearest-neighbor hopping amplitude and $U$ is the on-site Coulomb interaction. Here $c_{j,\sigma}^\dagger$ creates an electron with spin $\sigma=\uparrow,\downarrow$ on site $j$, and $n_{j,\sigma}=c_{j,\sigma}^\dagger c_{j,\sigma}$ is the corresponding number operator. We also define the total density and local spin operators as $n_j=n_{j,\uparrow}+n_{j,\downarrow}$ and $S_j^z=\frac{1}{2}(n_{j,\uparrow}-n_{j,\downarrow})$, respectively. The ground-state properties and the structure of spin and charge excitations in the one-dimensional Hubbard model are well established~\cite{text_Giamarchi,PRB.65.153110} and the polarization in a finite system has been investigated \cite{PRB.61.7883}. We use them as benchmarks in addition to the XXZ model discussed in the main text. In the numerical calculations, fermionic anticommutation relations are taken into account through the Jordan-Wigner transformation of the fermionic operators. We perform the iDMRG calculations with the particle-number conservation and total $S^z$ conservation.

We first present results at quarter filling, $\nu_{\mathrm{c}}=1/2$. In the zero-magnetization sector ($M=0$), one expects quantization with $q_{\mathrm{c}}=2$ for the charge sector and $q_{\mathrm{s}}=2$ for the spin sector. For repulsive interactions $U>0$, it is known that both of the charge and spin sectors are gapless, realizing a spin-density-wave (SDW) metallic state. In contrast, for attractive interactions $U<0$, the charge sector remains gapless while a gap opens in the spin sector, leading to a Luther--Emery liquid. Figure~\ref{fig:hubbard_1d}(a)(b) shows numerical results of $|z_{\mathrm{c}}^{2}|$ and $|z_{\mathrm{s}}^{2}|$. For $U>0$, both quantities approach zero, which is consistent with gapless charge and spin excitations. For $U<0$, $|z_{\mathrm{c}}^{2}|$ remains close to zero, whereas $|z_{\mathrm{s}}^{2}|$ increases toward unity, reflecting the opening of the spin gap in the Luther--Emery phase.

Next, we consider half filling, $\nu=1$. At half filling, for repulsive interactions $U>0$, the charge sector is gapped by the Umklapp scattering while the spin sector remains gapless. 
We expect quantization of the polarization amplitudes with $q_{\mathrm{s}}$ = 1 for the spin sector and $q_{\mathrm{c}}$ = 2 for the charge sector. Figure~\ref{fig:hubbard_1d}(c) presents $|z_{\mathrm{c}}|$ and $|z_{\mathrm{s}}^2|$ as functions of $L_{\mathrm{tw}}$. As $L_{\mathrm{tw}}$ increases, the spin polarization amplitude approaches zero, while the charge polarization amplitude tends to unity, in agreement with the known gap structure. 
These results demonstrate that, in iDMRG calculations, polarization-amplitude-based gap diagnostics are effective not only for pure spin systems but also for itinerant systems with charge degrees of freedom.

\section{Dependence on Bond-dimension}
\label{app:bond_dependence}
An important remark in iDMRG calculations is the dependence on the bond-dimension $D$. For an MPS with a finite $D$, one can define an effective correlation length, 
\begin{align}
    \xi_D \equiv \frac{1}{\mathrm{ln} \left( \frac{\lambda_2}{\lambda_1} \right) },
\end{align}
where $\lambda_1$ and $\lambda_2$ are the largest and second-largest eigenvalues of the transfer matrix\cite{SCHOLLWOCK201196}. A finite $\xi_D$ implies that the physical quantities behave as if the system had a finite correlation length, even when the true ground state is critical. Equivalently, there is an effective energy gap $\Delta_D \propto 1/ \xi_D$. 
Consequently, even if the true ground state is gapless, an MPS approximation with a finite $D$ necessarily exhibits a nonzero $\Delta_D$ and behaves effectively as a gapped system.

The correlation length $\xi_D$ directly affects the evaluation of polarization amplitudes based on twist operators. Since the twist operator acts on a region of length $L_{\mathrm{tw}}$, its expectation value can become dominated by the effective gap induced by finite $D$ once $L_{\mathrm{tw}}$ exceeds $\xi_D$. As a result, in a gapless phase where the polarization amplitude should approach $|z| \to 0 \, (|z_{\alpha}^{q_{\alpha}}| \to 0)$ with increasing $L_{\mathrm{tw}}$, one may instead observe a crossover toward gapped-like behavior for large $L_{\mathrm{tw}}$. This effect is clearly seen in the $L_{\mathrm{tw}}$ dependence of the spin polarization amplitude in the TLL of the XXZ chain, the XY phase of the XY--$J_{\chi}$ model, and the SDW phase of the Hubbard model (Fig.~\ref{fig:xi_dependice}). While the polarization amplitude first decreases monotonically and approaches zero as $L_{\mathrm{tw}}$ increases, beyond a characteristic scale $L_{\mathrm{tw}}^{\ast}$ it turns around and starts to increase. The numerical result for $L_{\mathrm{tw}}^{\ast}\sim \xi_D$ indicates that evaluating the polarization amplitude on regions larger than $\xi_D$ makes the result sensitive to the finite-$D$ effective gap $\Delta_D$, and therefore it does not faithfully reflect the true gap structure in the thermodynamic limit.
On the other hand, in the gapped regime, the value of $|z^q|\,(|z_{\alpha}^{q_{\alpha}}|)$ is essentially insensitive to further increase of $D$ once $\xi_D$ exceeds the physical correlation length.

These considerations imply that, when using polarization amplitude as a diagnostic of gapped versus gapless behavior, it is essential to (i) employ sufficiently large bond dimensions to ensure a large $\xi_D$, and (ii) choose the twist length within the regime $L_{\mathrm{tw}} \ll \xi_D$ and systematically check the convergence with respect to $L_{\mathrm{tw}}$ (and $D$).

\bibliographystyle{apsrev4-2}
\bibliography{ref}
\end{document}